\documentclass[10pt,prl,twocolumn]{revtex4}

\usepackage{graphicx}

\newcommand{\be}{\begin{equation}}
\newcommand{\ee}{\end{equation}}
\newcommand{\bea}{\begin{eqnarray}}
\newcommand{\eea}{\end{eqnarray}}

\newcommand{\f}{\frac}
\newcommand{\vdimer}{{\vrule height0.2cm width0.05cm depth0pt}}
\newcommand{\hdimer}{{\hrule height0.05cm width0.2cm depth0pt}}

\newcommand{\verdimers}{\hbox{\vdimer \hskip 0.1cm \vdimer}}
\newcommand{\hordimers}{\hbox{\vbox{\hdimer \vskip 0.1cm \hdimer}}}

\begin{document}

%\markboth{Olav F. Sylju{\aa}sen}{RANDOM WALKS NEAR ROKHSAR-KIVELSON POINTS}

%\title{RANDOM WALKS NEAR ROKHSAR-KIVELSON POINTS}
\title{Random walks near Rokhsar-Kivelson points}

\author{Olav  F. Sylju{\aa}sen}
\address{NORDITA, Blegdamsvej 17, DK-2100 Copenhagen {\O}, Denmark\\
sylju@nordita.dk}

%\revised{Day Month Year}
%\accepted{(Day Month Year)}
%\comby{(xxxxxxxxxx)}
%\end{history}

%\date{\today}

%\pacs{74.20.Mn, 75.10.Jm, 02.70.Ss}
%\preprint{NORDITA-2005-21}

\begin{abstract}
There is a class of quantum Hamiltonians known as Rokhsar-Kivelson(RK)-Hamiltonians for which static ground state properties can be obtained by evaluating thermal expectation values for classical models. The ground state of an RK-Hamiltonian is known explicitly, and its dynamical properties can be obtained by performing a classical Monte Carlo simulation. 
We discuss the details of a Diffusion Monte Carlo method that is a good tool for studying
statics and dynamics of {\em perturbed} RK-Hamiltonians without time discretization errors. 
As a general result we point out that the relation between the quantum dynamics and classical Monte Carlo simulations for RK-Hamiltonians follows from the known fact that the imaginary-time evolution operator that describes optimal importance sampling, in which the exact ground state is used as guiding function, is Markovian. Thus quantum dynamics can be studied by a classical Monte Carlo simulation for any Hamiltonian that is free of the sign problem provided its ground state is known explicitly.  
\end{abstract}

\maketitle

%\keywords{Rokhsar-Kivelson Hamiltonian, Continuous-time Diffusion Monte Carlo, quantum dimer model, plaquette phase.}

\section{Introduction}
Microscopic models that describe electronic behavior of materials are seldom exactly solvable.
Nevertheless it is often found that lots of insight can be gained by studying
a solvable model that resembles the accurate microscopic model in question.
A typical approach is to neglect certain couplings in the accurate microscopic model
in order to obtain a solvable model.
This brings up the question of how important the neglected couplings are.  One way of finding out is to simulate the accurate model numerically
and compare to the results of the solvable model.
If the results are sufficiently similar
one can, with reasonable confidence, claim to understand the physical behavior using
insights gotten from the solvable model.

The number of exactly solvable models are however relatively small. They tend
to be either models of non-interacting particles, like free fermions, or non-trivial,
but one-dimensional. Fortunately understanding and insight isn't necessarily
tied to exactly solvable models. Models that can be fully or partially
mapped onto other well-known models are also useful as older established results can be recycled.

There is a class of quantum models for which static ground state properties can be calculated
by evaluating thermal expectation values of a corresponding {\em classical} model
having the {\em same} number of space dimensions.
This is a tremendous simplification as much are known about classical statistical mechanics
models.

The most well-known of these quantum models is the quantum dimer
model(QDM) at a special point in parameter space; the Rokhsar-Kivelson(RK) point\cite{RK}.
There are also many other models with this property\cite{Henley,Ardonne,Castelnovo} 
with Hamiltonians known as (generalized) RK-Hamiltonians.
Beside static properties it is also possible to obtain information about excited states and dynamical properties for RK-Hamiltonians, although in a rather unorthodox way. Henley\cite{HenleyStat}
showed that dynamical correlation functions at the RK-point can be gotten by performing a continuous-time Monte Carlo simulation of the classical
model using appropriate Monte Carlo dynamics, and interpreting time-
correlation functions of the classical simulation as imaginary-time correlation functions of the quantum system.
This observation was recently utilized in Ref.~\cite{Ivanov} to determine
excitations of the quantum dimer model on the triangular lattice at the RK-point.
The observation that quantum dynamics can be gotten from classical
Monte Carlo simulations is however not peculiar to RK-Hamiltonians. It is far more general, but
requires knowledge of the {\em exact} ground state wave function.
This was recently pointed out in Ref.~\cite{Castelnovo}, but follows also, as will be
explained here, as a consequence using importance sampling with an optimal guiding function in Quantum Monte Carlo.

Assuming that the physics at the RK-point is understood or at least that it can be calculated
relatively easily, it is interesting to ask if the same physics also holds away from the RK-point.
An obvious approach to attempt answering this is to perform a finite-temperature path-integral Monte Carlo simulation of
the full quantum model. This is more complicated than performing a classical Monte Carlo simulation
at the RK-point, but can be carried out for most quantum models provided the temperature
is high enough. This approach was followed by Moessner and Sondhi in order
to show that the spin liquid state existing at the RK-point of the QDM on the triangular lattice
also extends to other values of the parameters\cite{MoessnerSondhi}.
However it is difficult to push these methods to low temperatures. The non-local update techniques usually
employed to speed up quantum Monte Carlo simulations such as loops\cite{Evertz}, worms\cite{Worm} or directed-loops\cite{SS}
are not easily applicable to the quantum dimer model as the configurations on each time-slice are
heavily constrained. The directed-loop method have however been used to study the classical dimer model,
for which very impressive system sizes can be studied effectively\cite{DimerSandvik}.
On the other hand the special Monte Carlo dynamics implied by the non-local moves doesn't 
correspond to the dynamics of the quantum dimer model at the RK-point.

In this review we will discuss another Monte Carlo technique which is an
excellent tool to study models in the vicinity of RK-Hamiltonians at zero temperature.
This technique, which is a continuous-time formulation of the well known
Diffusion Monte Carlo(DMC) technique, becomes equivalent to a continuous-time classical Monte
Carlo method for RK-Hamiltonians. In fact, as is true for any DMC method, it becomes equivalent to a classical Monte Carlo method whenever it is used in conjunction with importance sampling
that employs the exact ground state wave function as guiding function.

We concentrate on systems close to RK-points here. As is well known
with DMC-methods the quality of the results
depend to a large extent on the quality of the guiding function. The advantage
of being close to a RK-point is that the ground state wave function is known exactly at
the RK-point and can be used as good approximate guiding function in its vicinity. 

This review is divided into two main parts. In Sec.~\ref{Sec:QMC} the numerical method
is discussed in details. Particular emphasis is put on how to
extract observables reliably. In Sec.~\ref{Sec:QDM} the
application of the method to
perturbed RK-Hamiltonians is discussed, and results pertaining
to the quantum dimer model on a square lattice are given.

\section{Quantum Monte Carlo \label{Sec:QMC}}

There are two main branches of Quantum Monte Carlo methods.
The first type of methods uses a stochastic process to sample the finite temperature quantum partition function and extract observables from this. The main challenge using these methods is to engineer efficient updates as the objects, or variables, in the partition function are extended objects, or paths, as follows from the path integral formulation of quantum mechanics.
A big step forward in finding efficient update schemes has been achieved using avatars of the Swendsen-Wang\cite{SwendsenWang} cluster update, so called loop\cite{Evertz}, worm\cite{Worm} or directed-loop\cite{SS} updates. These Monte Carlo methods are exact in the sense that they have no systematic bias of any sort, they are easily formulated in continuous-time, and are easily programmable.  
A drawback, but also a source of flexibility, is the different ways one
can construct these non-local updates. This needs to be reconsidered for all models
in question, although it can be automated to a great extent and rules of thumb
for how to choose good rules exists\cite{Olav,Alet}.
  The finite-temperature methods are most efficient at high temperatures, and
the zero temperature behavior is only obtained asymptotically by performing simulations at decreasingly lower temperatures.

The other branch of Monte Carlo methods use stochasticity to
simulate the outcome of (repeated) matrix multiplications. This general class
of methods is known as Projector Monte Carlo(PMC), -repeated multiplications of the same matrix projects out the eigenstate with
the highest eigenvalue. For quantum systems one is interested in the {\em lowest}
eigenvalue of the Hamiltonian. 
Thus in Green function Monte Carlo which is a particular
projector Monte Carlo technique the iterated matrix is the inverse of the Hamiltonian. Another method, DMC, uses the matrix $\exp( -H \tau)$ where $H$ is the Hamiltonian and $\tau$ is imaginary-time.

For long times, or equivalently many matrix multiplications,
any initial state with some overlap with the true ground state will sample the components
of the ground state wave function and ground state observables can be obtained.
Moreover because DMC evolves the wave function identically
to the imaginary-time evolution quantum dynamic observables can be obtained
easily from the evolution of the state.

PMC techniques complement the finite-temperature techniques because they are genuine zero-temperature methods, -no extrapolations to zero temperature are necessary. They are also excellent at finding the ground state energy in different sectors of conserved quantum numbers, something which generally is difficult with the methods that work in the grand-canonical ensemble.
Moreover no ingenuity in finding efficient updates is needed as everything
is directly dictated by the Hamiltonian itself. 

The evolution matrix in DMC is in general not a Markovian matrix. Thus DMC
cannot be directly interpreted as a classical Monte Carlo method. The
standard way\cite{Donsker} to cope with the lack of probability
conservation is to include additional branching processes. Unfortunately
these have a tendency to make simulations unstable and some feedback
control is needed. This feedback results in a systematic bias\cite{Hetherington} to the observables which can be removed by reweighting the simulation at the expense of additional statistical errors. 

It is often stated repeatedly in the literature that DMC cannot be formulated in continuous (imaginary) time,
and that repeated runs with decreasingly smaller time-intervals must be performed in order to quantify the error induced by a finite time-step. However the construction and implementation of a continuous-time formulation
of DMC is straight-forward for lattice models\cite{OlavDimer}, a fact we aim to explain in the next section.
Other lattice PMC techniques\cite{Trivedi} using different operators to project onto the
ground state, such as for instance $1-(H-E_0)\tau$, are also free of time-discretization errors.
However these methods do not have the advantage, as do DMC, that quantum dynamics can be obtained easily. 
For another newly developed continuous-time Monte Carlo method, see Ref.~\cite{Schmidt}.

\subsection{Continuous-time Diffusion Monte Carlo algorithm \label{CTDMC}}

The basic idea behind DMC is to simulate the power
method stochastically.
The power method uses repeated matrix multiplications to project out the
eigenstate having the largest eigenvalue.
In DMC the imaginary-time evolution operator
projects out the ground state of the Hamiltonian. Specifically
\be
e^{-(H-E_0)\tau} |x_I \rangle \stackrel{\tau \to \infty}{=} C_{x_I} | \psi_0 \rangle
\ee
where $H$ is the Hamiltonian, $E_0$ is the ground state energy and  $|x_I\rangle$ is an arbitrary
initial state having overlap $C_{x_I}=\langle \psi_0 | x_I \rangle$ with the ground state
$| \psi_0 \rangle$.

We will now explain how the multiplication by $e^{-(H-E_0)\tau}$ is carried out in
continuous-time, that is without discretizing $\tau$.
Consider an N-state system with Hamiltonian
matrix elements: $H_{ij}= \langle i | H | j \rangle$, where $i,j = \{1,2,\ldots , N\}$.
For this $N$-state system an instance of the (unnormalized) wave function is described by an $N$-dimensional vector
having non-negative integer entries
\be
     | \psi \rangle = \left( \begin{array}{c} M_1 \\ \vdots \\ M_N \end{array} \right).
\ee
This integer-valued vector represents $M=M_1+M_2+\ldots+M_N$ replicas or copies of the system,
where $M_1$ of them are in state number 1, $M_2$ are in state 2, etc.. For big systems the dimension of
the vector is huge. With a finite number of replicas it will be very sparse and it is
better to keep track of the state of each replica than writing down the vector explicitly.
The requirement that the number of replicas in a given state is non-negative is rather restrictive,
and is equivalent to the requirement that there is no sign problem. We will restrict ourself
to these cases.

In order to build up the continuous-time formulation we will consider the evolution for an infinitesimal time step and then explain how to piece together (infinitely) many of these time steps in one shot.
The action of the time evolution operator for an infinitesimal time step $d \tau$ on an instance of the state is
\be \label{eq:evolution}
\left(
\begin{array}{c} M_1^\prime \\ M_2^\prime \\ \vdots \\ M_N^\prime \end{array} \right)
=
\left(
\begin{array}{cccc} D_{11} & -H_{12} d \tau & \cdots & -H_{1N} d\tau \\
                   -H_{21}d \tau & D_{22} & \cdots  & -H_{2N} d\tau \\
		   \vdots & \vdots &  & \vdots \\
	 	   -H_{N1}d \tau & -H_{N2} d\tau & \cdots & D_{NN}
\end{array} \right)
\left(
\begin{array}{c} M_1 \\ M_2 \\ \vdots \\ M_N \end{array} \right)
\ee
where the diagonal elements are $D_{ii} \equiv 1+(E_R-H_{ii}) d\tau$.
Note that the ground state energy, $E_0$, is not known at the outset of the simulation. Therefore an estimator
of the ground state energy known as the reference energy $E_R$ is introduced and used instead.
During the course of the simulation this reference energy will be adjusted and can be used to
extract the ground state energy. One should note that a time varying $E_R$ causes the evolved wave function to deviate from the ground state wave function\cite{Hetherington}.
This can be repaired by re-weighting the simulation as will be discussed in Sec.~\ref{Sec:Reweighting}.

We will now formulate a stochastic process that on average yields the
evolution equation, Eq.~(\ref{eq:evolution}).
In the time interval $d \tau$ a replica in
state $|i\rangle$ can undergo one out of four different processes with associated probabilities:
\begin{itemize}
\item
``Transition'', change state to $|j \rangle$, $j \neq i$, probability $P_{Tj}(i)$.
\item
``Die'', that is $M_i \to M_i-1$, probability $P_D(i)$.
\item
``Replicate'', that is $M_i \to M_i+1$, probability $P_R(i)$.
\item
``Stay'', stay unchanged in state $|i\rangle$, probability $P_S(i)$.
\end{itemize}
The ``Die'' and ``Replicate'' processes are known as {\em branching} processes.
As these are {\em all} possibilities, probability conservation implies
\be \label{eq:probconservation}
   P_{Tj}(i)+P_D(i)+P_R(i)+P_S(i) = 1,
\ee
and must hold for all states $i=1,\ldots, N$.

The task of identifying the probabilities with matrix elements of the
Hamiltonian is easy.
Because the off-diagonal matrix element $H_{ji}$ is the only one responsible
for transition between state $i$ and $j$ it is clear that
\be \label{eq:transition}
    P_{Tj}(i) = -H_{ji} d\tau
\ee
where $j \neq i$. In order to avoid the sign problem off-diagonal
matrix elements are restricted to be negative.

The increase in the number of replicas in state $|i\rangle$ from processes acting on replicas in state $|i\rangle$ is
\be \label{eq:changeindiagreplicas}
  M_i^\prime - M_i = \left[ P_R(i)-P_D(i)- \sum_{j \neq i} P_{Tj}(i) \right] M_i.
\ee
This implies when comparing to the diagonal elements of Eq.~(\ref{eq:evolution}) and using Eq.~(\ref{eq:transition})
that
\be \label{eq:replicatingdying}
   P_D(i)-P_R(i) = \left(H_{ii}-E_R+ \sum_{j \neq i} H_{ji} \right) d \tau.
\ee
The right hand side of the above takes either a positive or a negative value.
In order to reduce the fluctuations in the replica numbers as much as possible
$P_R=0$ is chosen whenever this value is positive and $P_D=0$ when it is negative.
This choice implies that $P_D$ and $P_R$ are of the order $d \tau$ as also holds for $P_T$.
The probability conservation equation Eq.~(\ref{eq:probconservation}) then implies
\be \label{eq:staying}
  P_S(i) = 1 -  \left( |  H_{ii}-E_R+\sum_{j \neq i} H_{ji} | - \sum_{j \neq i} H_{ji} \right)  d\tau,
\ee
which is of the order unity.

We are now at the stage where we wish to patch together many infinitesimal
time steps. From the fact that $P_S$ is of the order unity and all other processes
are of the order $d\tau$ it follows that for most
(infinitesimal) time intervals nothing happens to a replica.
The process can thus be modeled like
the radioactive decay problem, although with several different decay channels: ``Transition'',``Die'', and
``Replicate''. The ``Transition'' channel is further divided into possibly as many as $(N-1)$ different channels
corresponding to different non-zero values of $H_{ji}$.
This observation has been used previously to construct continuous-time
algorithms for finite-temperature quantum Monte Carlo methods \cite{BeardWiese}.

\begin{figure}
\begin{center}
\includegraphics[clip,width=6cm]{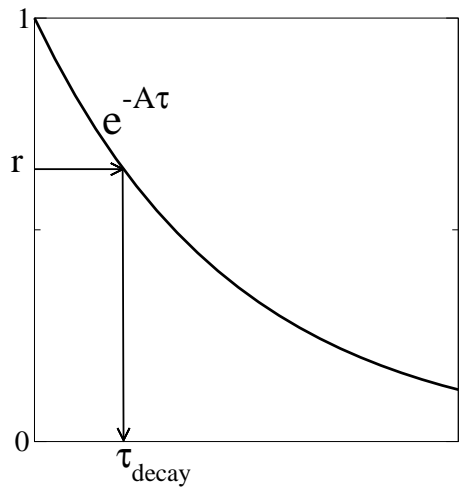}
\caption{
Selecting decay times according to the exponential distribution with decay constant $A$.
Drawing a random number $r$, the decay time $\tau_{\rm decay}$ is selected
as $\tau_{\rm decay} = \f{1}{-A} \ln(r)$. \label{simexp}}
\end{center}
\end{figure}

It follows that the imaginary-time evolution of one replica can be simulated by generating exponentially
distributed decay times with decay constant $A=|H_{ii}-E_R+\sum_{j \neq i} H_{ji}|- \sum_{j \neq i} H_{ji}$, see Fig.~\ref{simexp}.

Having obtained the decay time, the type of decay is determined
stochastically proportional to the respective probabilities $P_T, P_D$ and $P_R$, which are
all of the same order $d\tau$.

\subsection{Practical implementation \label{Sec:PractImp}}
Although the method is formulated in continuous-time it is convenient to use
discrete control times $\tau_c^i$ at regularly spaced time-intervals $\Delta \tau$ as times where
measurements are recorded and population control are being performed.

\begin{figure}
\begin{center}
\includegraphics[clip,width=8cm]{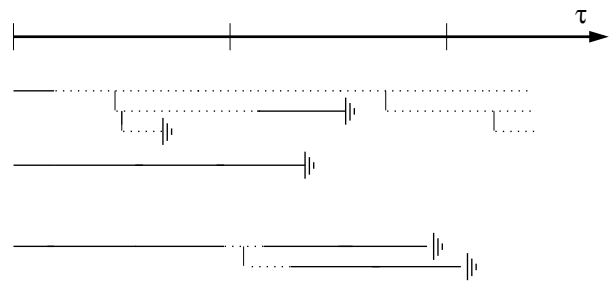}
\caption{
Illustration of a possible evolution for a 2-state system. The top line is the imaginary-time axis, on which
control times are indicated by vertical bars. Initially there are three replicas, all in state 1 labeled
by solid lines. As time progresses replica 1 first ``Transitions'' to state 2 (dotted line),
and then ``Replicates''. Its replica also ``Replicates'', but its copy ``Dies'' (the grounding symbol) almost immediately. Thus at the first control time
the first replica has changed state and has divided into two. Nothing happens to the second replica up to the first control time, it just stays in the state 1. The third replica ``Transitions'' to state 2 just before the first control time. Having propagated all three initial replicas and their descendants  up to the first control time, the population control procedure is performed before
propagation up to the next control time starts.\label{practvis}}
\end{center}
\end{figure}

In an actual simulation each replica contains information about the state of the system as well as a ``clock''
indicating the starting time for the next evolution. The replicas are ordered in a list.
They all start in the same state and with their clock set to 0.
Each replica in the list is subsequently evolved up to a control time $\tau_c$,
or until the replica dies in which case it is removed from the list. An evolution of a replica
begins by generating a decay-time $\tau_d$ according to the exponential distribution. If
$\tau_d > \tau_c$ the clock is set to $\tau_c$ and evolution of the next replica in the list starts.
If however $\tau_d < \tau_c$, the clock is set to $\tau_d$ and a random number is drawn to
select the decay type. If the decay type is ``Transition'' the state of the replica changes, if
it is ``Die'' the replica is removed from the list, and finally 
the decay-type ``Replicate'' causes a copy of the replica with clock set to $\tau_d$ 
to be inserted at the end of the list.
As long as the replica is not dead the evolution continues by picking a new decay time until $\tau_c$ is reached.
When the last replica in the list has evolved up to $\tau_c$, all replicas have the same clock-time, and
measurements can be performed. The process is repeated by increasing $\tau_c$ and starting over from the beginning of the replica list. A graphical visualization of a possible evolution of replicas is shown in
Fig.~\ref{practvis}.

The control times are included in order to perform population control to avoid an explosion/implosion in the number of replicas. Population control is achieved by changing the value of the reference energy $E_R$ so as to maintain a roughly constant number of replicas. $E_R$ is only adjusted at the control times and
is kept constant in between.
Specifically, with $E_R^{i}$ denoting the reference energy just {\em after} control time $\tau_c^{i}$, which
is the same value as just before the next control time $\tau_c^{i+1}$, a possible choice for population control is
\be
	E_R^{i} = \overline{E_R} + \f{1}{\Delta \tau} \ln \left( \f{N_0}{N_i} \right)
\ee
where $N_i$ is the total number of replicas at $\tau_c^i$ and $N_0$ is the total
number of replicas at the beginning of the simulation. $\overline{E_R}$ denotes the average value
of $E_R$ for all control time points up to $\tau_c^i$.
Thus when the number of replicas decreases below $N_0$ the reference energy is raised which
will tend to increase the replication process, thereby increasing the replica population.

\subsection{Observables}
The measurement of observables in DMC requires some care\cite{NightingaleUmrigar}.
There are two issues which need to be addressed. First one must get rid
of the dependence on the choice of initial and final wave function.
Then one must be careful about getting rid of the bias introduced by the population control procedure.  Lets first explain how to get rid of the dependence on the initial and final wave functions.  This is known as the  forward- or future-walking method\cite{forward}.

When the initial wave function is a particular basis state
$|x_I \rangle$ the power method yields
\be
	e^{-H\tau} | x_I  \rangle \stackrel{\tau \to \infty}{=} C_{x_I} | \psi_0 \rangle
\ee
where the overlap $C_{x_I}=\langle x_I | \psi_0 \rangle$. For simplicity in notation we
have absorbed the reference energy into the Hamiltonian in this section.
The projection can of course also be carried out for a wave function which is the superposition
(with unit coefficients) of all orthonormal basis states $| 1 \rangle \equiv \sum_x | x \rangle$
\be
	e^{-H\tau} | 1 \rangle \stackrel{\tau \to \infty}{=} C_1 | \psi_0 \rangle.
\ee
This can be converted into an evolution of the dual state
\be
	\langle 1| e^{-H\tau} \stackrel{\tau \to \infty}{=} \langle \psi_0 | C_1.
\ee
Note that the overlaps are all assumed to be real and positive (in accordance with
the restriction of having no sign problem).

In DMC the matrix evolution $e^{-H\tau} |x_I \rangle$ is replaced by a stochastic process
\be
	e^{-H\tau} | x_I \rangle = \sum_x P(x,\tau ; x_I,0) |x \rangle
\ee
where $P(x,\tau;x_I,0)$ is the probability of finding the evolved state in the basis state $ |x\rangle$ at time $\tau$ provided it was in state $|x_I\rangle$ at time zero.
The stationary distribution of this evolution is proportional to the ground state. It follows
that the following can be used as an estimator of the ground state
\bea
	| \psi_0 \rangle & \stackrel{N_\tau \to \infty}{=} & 
	                   \f{1}{C_{x_I} N_\tau} \sum_{\tau} | \psi_\tau \rangle \nonumber \\
		& = & \f{1}{C_{x_I} N_\tau} \sum_{\tau} \sum_x P(x,\tau; x_I,0) | x \rangle
\eea
where the sum over different values of $\tau$ contains $N_\tau$ terms and the first $\tau$-value
is taken after an initial equilibration value $\tau_e$.

Using these relations one can obtain an expression for the ground-state matrix element of
the observable described by the operator  ${\cal O}$
\be
	\langle \psi_0 | {\cal O} | \psi_0 \rangle \stackrel{\tau^\prime,N_\tau \to \infty}{=}
	\alpha \sum_{\tau} \langle 1 | e^{-H \tau^\prime} {\cal O} | x \rangle P(x,\tau; x_I,0)
\ee
as well as for the norm of the ground state
\be
	\langle \psi_0 | \psi_O \rangle \stackrel{\tau^\prime,N_\tau \to \infty}{=}
	\alpha \sum_{\tau} \langle 1 | e^{-H \tau^\prime} | x \rangle P(x,\tau; x_I,0).
\ee
To shorten notation we have collected the overlap coefficients and the number
of measurements into a single constant $\alpha=1/(C_{x_I} C_1 N_{\tau})$.
The overlap coefficients cancel when considering the ratio
\be
	\langle {\cal O} \rangle = \f{\langle \psi_0 | {\cal O} | \psi_0 \rangle}{\langle \psi_0 | \psi_0 \rangle}.
\ee

\subsubsection{Diagonal observables}
Lets now specialize to the case where the observable ${\cal O}$ is a time-independent observable diagonal in the basis set $\left\{ |x\rangle \right\}$. Then the ground state matrix element is
\bea
\lefteqn{	\langle \psi_0 | {\cal O } | \psi_0 \rangle } \nonumber \\
& = & \alpha \sum_{\tau} \sum_{x,x^\prime} \langle 1 | x^\prime \rangle \langle x^\prime | e^{-H \tau^\prime}
	 {\cal O} |x \rangle P(x,\tau ; x_I,0)
	\nonumber \\
	& = & \alpha
	\sum_{\tau} \sum_{x,x^\prime} \langle 1 | x^\prime \rangle P(x^\prime,\tau+\tau^\prime; x, \tau)
	{\cal O}_{x}  P(x,\tau; x_I,0) \nonumber \\
	& = & \alpha \sum_{\tau} \sum_{x,x^\prime} P(x^\prime,\tau+\tau^\prime; x, \tau)
	{\cal O}_{x}  P(x,\tau; x_I,0)
\eea
where we have used $\langle 1 | \equiv  \sum_{x^{\prime \prime}} \langle x^{\prime \prime} | $, orthonormality of the set of basis states and the eigenvalue relation $ {\cal O} | x \rangle= |x \rangle {\cal O}_{x}$. Similarly the norm is
\be
	\langle \psi_0 | \psi_0 \rangle = \alpha \sum_\tau \sum_{x^\prime} P(x^\prime, \tau+\tau^\prime; x_I,0)
\ee
In both these expressions the limits $\tau,\tau^\prime \to \infty$ are implied.
This limit can of course not be achieved in practice so instead one evaluates the right hand sides with large values of $\tau$ and $\tau^\prime$.

Starting with typically thousands of replicas in the same initial state $|x_I\rangle$ the probability
$P(x,\tau; x_I,0)$ is proportional to the number of replicas in state $|x\rangle$ at time $\tau$.
Thus the denominator is simply the total number of replicas at time $\tau+\tau^\prime$. The numerator
is a bit trickier in that one counts the replicas at time $\tau+\tau^\prime$ but with each replica weighted by the value of the observable at time $\tau$. To implement this in practice it is convenient to let
each replica keep track of its history of observation values. This list must at least be of the length $\tau^\prime$.
Alternatively one could let each replica keep a history of its configurations, but this is quite
memory-consuming, and it is better to concentrate on a few observables and their histories.
The history list is such that whenever a walker replicates, the history list is also
replicated such that the new walker inherits the same history as the original walker.

The measurements must be performed at identical times for all replicas.
One way of doing this is to measure the observable at the control times and
store the result in the history list for each replica. One can also pick the measurement
point to be at an arbitrary time in the time interval between replicas, instead
of at precisely the control time itself. In fact, one can take the average of
the observable over all time points in the interval, so as to get the maximum
information available.
This is quite easy to implement as the observable only changes value when a
decay of the ``Transition'' type happens. As the times of decays are known it is quite
easy to compute the time-weighted average of the observable for the interval
between control times.
 For instance, assuming that there is just one decay at $\tau_d$ changing the observable
from ${\cal O}_1$ to ${\cal O}_2$, the average value accumulated at the control time $\tau_c^i$ is then $({\cal O}_1 (\tau_d-\tau_c^{i-1})
+{\cal O}_2 (\tau_c^i-\tau_d))/(\Delta \tau)$. This average value is then stored in the replica's history
list of measurements.

\subsubsection{Off-diagonal observables}
It is slightly more complicated to measure off-diagonal static (time-independent) observables than diagonal observables as there
are no ways to insert off-diagonal operators without disturbing the configurations. The trick usually employed, which only works when the off-diagonal
operator is a term in the Hamiltonian is best appreciated by visualizing the
stochastic process as divided into small discrete time steps.
That is
\be
	P(x,\tau; x_I,0) =
	P(x,\tau; x_{n-1},\tau_{n-1}) \ldots P(x_2,\tau_2;x_1,\tau_1) P(x_1,\tau_1;x_I,0)
\ee
This discretization is never used in practice, it is used here merely for explaining the measurement procedure.
In the limit of infinitesimal time intervals the processes involving changes of states
are related to off-diagonal matrix elements of the Hamiltonian $P(y,\tau+d\tau;x,\tau)= -\langle y|H |x \rangle d\tau$, see Eq.~(\ref{eq:transition}).
The matrix element of the off-diagonal operator can then be written
\bea
\langle \psi_0 | {\cal O} | \psi_0 \rangle & = &\sum_{x,x^\prime,x^{\prime \prime}} P(x^\prime, \tau^\prime+\tau; x^{\prime \prime} \tau)
\langle x^{\prime \prime} | {\cal O} | x \rangle \nonumber \\
&  & \times P(x , \tau-d\tau; x_I,0)
\eea
Note the ``hole'' introduced, there is nothing between $\tau-d\tau$ and $\tau$. There
is nothing wrong with having such a hole as the probabilities are invariant under time translations. Thus we can think of the left-most factor as being time-translated by an amount $d \tau$.
Multiplying by 
\be
	 1 = \f{ P_{\cal O} (x^{\prime \prime}, \tau; x, \tau-d\tau)}
	 	{-\langle x^{\prime \prime} | H_{\cal O} | x \rangle d\tau}
\ee
one gets
\bea \label{eq:offdiagonal}
\lefteqn{
\langle \psi_0 | {\cal O} | \psi_0 \rangle =  \sum_{x,x^\prime,x^{\prime \prime}} P(x^\prime, \tau^\prime+\tau; x^{\prime \prime}, \tau)} \nonumber \\
&  & \times 
   P_{{\cal O}}(x^{\prime \prime},\tau; x,\tau-d\tau) P(x, \tau-d\tau; x_I,0) 
  \f{\langle x^{\prime \prime} | {\cal O} | x \rangle}
    {-\langle x^{\prime \prime} | H_{\cal O} | x \rangle d\tau} \nonumber \\
\eea
where the subscript ${\cal O}$ indicates the transition described by the observable ${\cal O}$.
There is nothing special about the time $\tau$, other than it should be far from the starting and final time, thus we might take the average over many time intervals each of length $d\tau$ within a range of nearby control times $(\tau_c^{i-1},\tau_c^i)$. Implementing this in practice amounts to counting the number of transitions corresponding to the off-diagonal operator in question within the time interval $\Delta \tau$ and multiplying this number by
$ \f{\langle x^{\prime \prime} | {\cal O} | x^\prime \rangle}{-\langle x^{\prime \prime} | H | x^\prime \rangle \Delta \tau}$.
This number is then stored as the measurement value at time $\tau_c^i$ in the history list, and
the variables are accumulated in the same way as for diagonal observables using the
forward-walking method where replicas are counted at $\tau+\tau^\prime$ weighted by
the measurement value at $\tau$.

\subsection{Dynamic observables}
Dynamic observables can be recorded from the history list of measurement results for each replica.
Dynamic correlation functions at times $\tau = m \Delta \tau$ significantly larger than the time
interval between control times $\Delta \tau$ can simply be gotten by taking products
of entries that differ by $m$ slots in the history list. In order to measure the small time
behavior using this approach $\Delta \tau$ should be made smaller. Of course, as the
method is formulated in continuous-time, there is nothing in principle restricting the measurement
times to discrete time points $m \Delta \tau$. At the expense of some extra book-keeping one can
store and record observables at any time-separations.

\subsection{Reweighting \label{Sec:Reweighting}}

The use of population control where the reference energy $E_R$ is changed in order to keep
the number of replicas approximately constant implies that the Monte Carlo procedure
with population control is not sampling exactly the imaginary-time evolution of the
wave function. Instead it is sampling an evolution with a time-dependent reference
energy. Varying $E_R$ according to the recipe described in Sec.~\ref{Sec:PractImp}
the resulting wave function $\overline{\psi}$ will differ from the correct ground state wave function
$\psi$ by a product of time-dependent factors\cite{Umrigar}

\be
	\psi(\tau_i) = e^{- \sum_{k=0}^{i-1} (E_R^k-E_R^0) \Delta \tau} \overline{ \psi} (\tau_i)
\ee
The subtraction of the constant $E_R^0$ is done in order to keep the exponential from overflowing.

The extra multiplicative factor coming from the time-varying $E_R$ can be gotten rid of
by multiplying by the exponential factor above, giving for a diagonal 
observable
\begin{widetext}
\be
	\f{\langle \psi_0 | {\cal O } | \psi_0 \rangle}{ \langle \psi_0 | \psi_0 \rangle} =
	\f{\sum_{x,x^\prime} P(x^\prime,\tau_i; x, \tau)
	{\cal O}_{x}  P(x,\tau ; x_I,0) e^{-\sum_{k=0}^{i-1} (E_R^k-E_R^0) \Delta \tau}}
	{\sum_{x} P(x^\prime, \tau_i; x_I,0) e^{-\sum_{k=0}^{i-1} (E_R^k-E_R^0) \Delta \tau } }
\ee
\end{widetext}
Of course one cannot keep track of an infinite product of factors, and if one could,
the fluctuations would be enormous, making it impossible to get accurate results\cite{Hetherington}.
However if the number of replicas isn't varying a lot, which can be achieved
using importance sampling as described in the next section, the fluctuations need not be big\cite{Umrigar}.

\begin{figure}
\begin{center}
\includegraphics[clip,width=7cm]{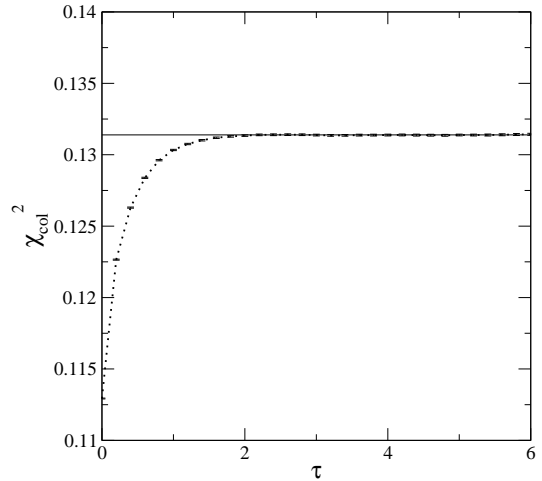}
\caption{
Columnar order parameter $\chi_{\rm col}^2$ on a 4x4 lattice vs. reweighting times
for the quantum dimer model at $V/J=0.1$. $M=1000$ replicas were used. 
The reweighting times are here defined using projections for $\tau$ steps to project onto the ground state and forward-walking for $\tau$ additional time steps.
The line is the exact diagonalization result.\label{columnar4x4}}
\end{center}
\end{figure}

In practice the observables are studied for different values of reweighting times.
Typically one sees a clear bias without reweighting which is possible to
avoid by taking longer reweighting times, however too long reweighting times gives
added noise. Fortunately the bias vanishes usually before the noise gets big, thus there
is a region where one can get accurate measurements, see Fig.~\ref{columnar4x4}.

\subsection{Importance Sampling \label{Sec:ImportanceSampling}}
It is known that importance sampling reduces statistical errors in DMC\cite{ImpSampling}. Importance
sampling is achieved by sampling the product of the wave function times a guiding function
instead of the wave function alone. The guiding function is chosen to be as close to the 
exact ground state as possible.
Lets now show that branching can be avoided using importance sampling when the 
guiding function coincides with the exact ground state wave function.
Consider the (infinitesimal time) evolution equation
\be
\langle \sigma | \psi(\tau + d\tau) \rangle = \sum_{\sigma^\prime}
	\langle \sigma | \left( 1 + (E_R-H ) d \tau \right) | \sigma^\prime \rangle
	\langle \sigma^\prime | \psi(\tau) \rangle
\ee
where $| \psi(\tau) \rangle$ labels the simulated instance of the wave function and  $|\sigma \rangle$ denotes a basis state.
Multiply by the time-independent guiding function $\langle \psi_g | \sigma \rangle$ and insert the unity factor $\f{\langle \psi_g | \sigma^\prime \rangle}{\langle \psi_g | \sigma^\prime \rangle}$
\bea
\lefteqn{ \langle \psi_g | \sigma \rangle \langle \sigma | \psi(\tau + d\tau) \rangle = } \nonumber \\
&  &\langle \psi_g  | \sigma \rangle  \sum_{\sigma^\prime}
	\langle \sigma | \left( 1 + (E_R-H ) d \tau \right) | \sigma^\prime \rangle
	 \f{\langle \psi_g | \sigma^\prime \rangle}{\langle \psi_g | \sigma^\prime \rangle}
	 \langle \sigma^\prime | \psi(\tau) \rangle. \nonumber \\
\eea
As all eigenfunctions are real here, we do not distinguish between $\langle \psi | \sigma \rangle$
and $\langle \sigma | \psi \rangle$.
The evolution matrix governing the evolution of the product $\langle \psi_g | \sigma \rangle \langle \sigma | \psi(\tau) \rangle$ is therefore
\be \label{modifiedevolution}
\langle \psi_g | \sigma \rangle
	\langle \sigma | \left( 1 + (E_R-H ) d \tau \right) | \sigma^\prime \rangle
	 \f{1}{\langle \psi_g | \sigma^\prime \rangle}.
\ee
Now take the case where the guiding function $\psi_g$ coincides with the true ground state wave function. Then when
we sum over columns of this matrix, that is $\sum_{\sigma}$ and use the resolution of the identity, $1=\sum_{\sigma} | \sigma \rangle \langle \sigma |$, we get
\bea
        \sum_{\sigma} 
        \langle \psi_g | \sigma \rangle
	\langle \sigma | ( 1  + \lefteqn{ (E_R-H ) d \tau ) | 
	                 \sigma^\prime \rangle 
	 \f{1}{\langle \psi_g | \sigma^\prime \rangle} }\nonumber \\ 
	 & = &
	 1+
	 \f{\langle \psi_g | \left( E_R - E_0 \right) d\tau | 
	 \sigma^\prime \rangle }{\langle \psi_g | \sigma^\prime \rangle}
	 \nonumber \\
\eea
The last term vanishes provided $E_R=E_0$.
The fact that each column sum to unity means that the matrix is Markovian.
The evolution can then be simulated entirely without branching processes 
and is thus
a {\em classical} Monte Carlo simulation in continuous-time.

The diagonal elements of the evolution matrix are not affected by the guiding function, while the off-diagonal ones are. This changes the transitions between the states. Basically
the transition to a state is enhanced if the guiding function has a large value
for that state. If the state it was coming from also corresponds to a large guiding function value, $\psi_g^{-1}$ ensures
that this is compensated for.

\section{Quantum dimer model on the square lattice \label{Sec:QDM}}
We will now apply the continuous-time DMC method to perturbed RK-Hamiltonians 
taking the QDM on a square lattice as an example.
The QDM was first proposed in the context of resonant valence bond (RVB) theories of high-Tc superconductivity\cite{KRS}. In the RVB theory\cite{Fazekas,AndersonRVB} pairs of spins form singlets that are approximated by dimers in the QDM.
The Hamiltonian is
\be \label{Hamiltonian}
	H= -J \sum \left(  \vphantom{\sum} | \verdimers \rangle  \langle \hordimers | + \rm{H.c.} \right)
	   +V \sum \left( \vphantom{\sum} | \verdimers \rangle \langle \verdimers | +
                                             | \hordimers \rangle \langle \hordimers |\right)
\ee
where the summations are taken over all elementary plaquettes of the lattice.
The basis states of the QDM consist of all possible dimer coverings of the lattice such
that all sites are covered and no dimers overlap each other.
The potential energy term $V$ counts the number of plaquettes possessing parallel dimers
and the $J$-term associates a kinetic energy to flips in the orientation of these parallel dimers.
For this reason plaquettes possessing parallel dimers are named flipable plaquettes.

Rokhsar and Kivelson\cite{RK} realized that one could find the exact ground-state of this model
at $J=V$. The ground-state at this RK-point is simply
the equal superposition of all basis states
\be \label{EqualSuper}
	|0 \rangle = \f{1}{\sqrt{N}} \sum_{i} | i \rangle
\ee
where $|i \rangle$ is a basis state describing a particular dimer covering of the lattice. $N$ is the total number of basis states.
Any ground-state expectation value of a time-independent operator diagonal in the basis states can then be evaluated as
\be \label{ThermalExp}
	\langle 0 | {\cal O} | 0 \rangle = \f{1}{N} \sum_i {\cal O}_i
\ee
where the sum goes over all classical dimer covering states. The right hand side can be
recognized as an expectation value in classical statistical mechanics taken at infinite temperature,
where the classical partition function becomes $Z=N$.

The special property of the QDM at $V=J$ which makes the simple
wave function Eq.~(\ref{EqualSuper}) the ground state is the fact that the Hamiltonian can be written as a sum weighted
by positive coefficients of
positive semi-definite Hermitean operators acting on pairs of states, where each operator is proportional to its own square.

Eq.~(\ref{ThermalExp}) describes a classical system at infinite temperature. The recipe for constructing
finite-temperature RK-Hamiltonians was given in Ref.~\cite{Castelnovo}.
A quantum Hamiltonian of the form, which one can view as the general defining form of RK-Hamiltonians,
\bea \label{HRK}
	H_{RK} & = & \sum_{\left\{ s,s^\prime \right\} } t_{s,s^\prime} \left(
	e^{-K(E_{s^\prime}-E_s)/2} |s\rangle \langle s| 
	\right. \nonumber \\
	& & \left. +e^{K(E_{s^\prime}-E_s)/2} |s^\prime \rangle \langle s^\prime| - |s\rangle \langle s^\prime | - |s^\prime \rangle \langle s| \right)
\nonumber \\
\eea
has the ground state
\be \label{HRKgs}
|\psi_0 \rangle = \f{1}{\sqrt{Z}} \sum_s e^{-K E_s/2} |s \rangle
\ee
with $E_0=0$, where $Z=\sum_s e^{-K E_s}$ and $t_{s,s^\prime}$ res positive.
It follows that the ground state expectation value of static quantities can be gotten
by evaluating a classical thermal expectation value at inverse temperature $K$.

\subsection{Quantum dynamics from classical Monte Carlo}

In addition to the ``classical'' ground-state properties of RK-Hamiltonians it was
pointed out by Henley that the quantum dynamics of the RK-Hamiltonian can
be gotten by performing a classical Monte Carlo simulation with the appropriate dynamics\cite{HenleyStat,Henley}. We want to show that this relation between quantum dynamics and classical Monte Carlo is general.
This was also recently discussed in Ref.~\cite{Castelnovo}. 
Before doing so lets see what happens to the continuous-time DMC method for the QDM exactly at the RK-point as
well as for the generalized RK-Hamiltonians in Eq.~(\ref{HRK}).

A state with $f$ flipable plaquettes in the QDM has potential energy $fV$.
Having $f$ flipable plaquettes implies that $f$ other states are accessible by one dimer flip, 
thus there is a kinetic energy $-fJ$ associated with this state. 
At the RK-point, $V=J$, the sum of potential and kinetic energy is thus zero. 
It follows from Eq.~(\ref{eq:replicatingdying}) that
$P_D-P_R=0$ provided $E_R$ is set equal to the true ground state
energy $E_0=0$.
This holds for any basis state of the system.  
The fact that branching can be set to zero means
that the DMC has been reduced to a classical Monte Carlo procedure. 

For the generalized RK-Hamiltonians described in Eq.~(\ref{HRK}) , the branching processes
are not automatically zero. However branching can be avoided if
one simulates the {\em product} of the wave function with the ground state wave function Eq.~(\ref{HRKgs}).  According to Eq.~(\ref{modifiedevolution}) the Hamiltonian then changes effectively,
 $H \rightarrow \psi_g H \psi_g^{-1}$, meaning that diagonal terms are unchanged while
off-diagonal terms, $H_{s^\prime s} \to e^{-K(E_{s^\prime}-E_s)/2} H_{s^\prime s}$, where
we have used the ground state of the RK-Hamiltonian Eq.~(\ref{HRKgs}) as guiding function.
The amount of branching needed according to Eq.~(\ref{eq:replicatingdying}) becomes again
\bea
P_D(s)-P_R(s) & = & \sum_{s^\prime \neq s} t_{s,s^\prime} e^{-K(E_{s^\prime}-E_s)/2} \nonumber \\
	& &	- \sum_{s^\prime \neq s} t_{s,s^\prime} e^{-K(E_{s^\prime}-E_s)/2} \nonumber \\
& = & 0, 
\eea
provided $E_R=E_0=0$.
Thus the DMC algorithm from which quantum dynamics can be extracted reduces
to a pure {\em classical} Monte Carlo simulation when used in combination with importance sampling
using the exact ground state as guiding function. 

It is peculiar to note that in the quantum dimer model case the
DMC method reduced to a classical Monte Carlo method without the explicit mentioning of a guiding function as was needed for the general RK-Hamiltonians. 
However, in fact the guiding function was used implicitly. This becomes clear when realizing that the ground-state Eq.~(\ref{EqualSuper}) is the equal amplitude superposition of all basis states, thus $\psi_g(s) \psi_g^{-1}(s^\prime) = 1$
for all states $s$ and $s^\prime$. In other words the equal amplitude wave function is the default guiding function used in DMC when no other explicit guiding functions are specified. It corresponds simply to the multiplication by unity.

Having explained how optimal importance sampling leads to a classical Monte Carlo procedure for RK-Hamiltonians it should be clear that the underlying reason follows from the known importance sampling argument originally presented in Ref.~\cite{ImpSampling} and restated in Sec.~(\ref{Sec:ImportanceSampling}); the operator governing the evolution 
of the wave function {\em times} the exact ground state wave function is a Markovian matrix. Thus quantum dynamics can be gotten from classical Monte Carlo simulations for {\em any} Hamiltonian
that can be simulated with the DMC method provided the exact ground-state is explicitly known. 
The advantage of dealing with RK-Hamiltonians is the fact that their ground states and energies are known explicitly from the construction of the models.

\subsection{Measurement results}
Now turn to the physics of the QDM on the square lattice.
For negative values of $V$ it is favorable to have parallel arrangement of dimers around
a plaquette. For positive $V$ there is a potential energy cost to have such parallel arrangements,
however the resonance term still favors parallel dimers. Thus it can be expected that
the favorable dimer configurations maximize the number of parallel dimers also for positive
$V$. In order to measure the dominance of parallel dimers we use the so called columnar
order parameter which can be written as
\be
\chi_{\rm col}^2 = \f{1}{4N^2} \langle
\left( \sum n_{\rm H}(\vec{r}) (-1)^{r_x} \right)^2
+
\left( \sum n_{\rm V}(\vec{r}) (-1)^{r_y} \right)^2
\rangle
\ee
where $n_{\rm H}$ ($n_{\rm V}$) is the number of horizontal (vertical) dimers belonging to the plaquette at $\vec{r}$. $\vec{r}$ is an integer-valued coordinate labeling the center of each plaquette on a lattice of size $N=L \times L$. The sums are to be taken over all plaquettes.

The columnar order parameter is a diagonal observable which does not commute with the Hamiltonian. It is
therefore necessary to use the forward walking technique, which require reweighting.

\begin{figure}
\begin{center}
\includegraphics[clip,width=7cm]{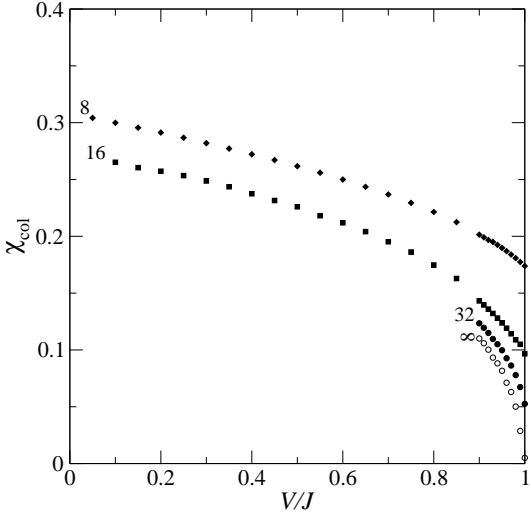}
\caption{
Columnar order parameter $\chi_{\rm col}$ vs. $V/J$.
The different curves are for different linear system sizes $L=8,16,32$ and the open
symbols are (quadratic) extrapolations to $L=\infty$.
\label{columnar}}
\end{center}
\end{figure}

Measuring the columnar order parameter for different system sizes at different values of $V/J$
we obtain the results shown in Fig.~\ref{columnar} \cite{OlavDimer}.
It is seen that the columnar order parameter remains finite for $V/J<1$ and goes to
zero at $V/J=1$. These results are consistent with earlier results obtained using
exact diagonalization\cite{Sachdev,Runge}.

\begin{figure}
\begin{center}
\includegraphics[clip,width=7cm]{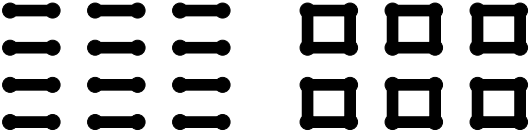}
\caption{
Columnar phase (left) and plaquette phase (right). The presence of bonds indicates bigger probabilities for dimers than the absence of bonds.
\label{colplaqfig}}
\end{center}
\end{figure}

The columnar order parameter does not distinguish between columnar and plaquette
phases, see Fig.~\ref{colplaqfig}. 

\begin{figure}
\begin{center}
\includegraphics[clip,width=7cm]{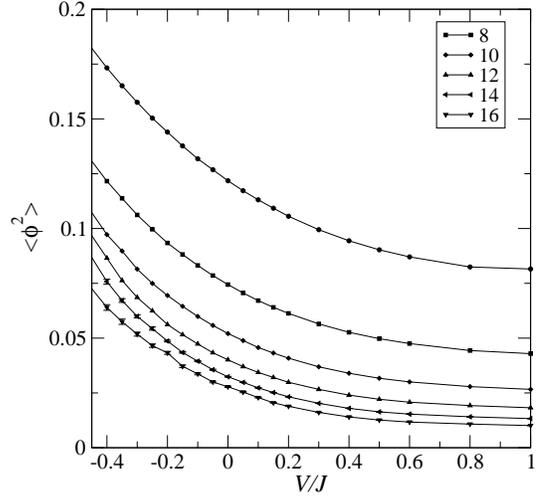}
\caption{
The value of $\langle \phi^2 \rangle$ vs. $V/J$ for different system sizes.
\label{phi2}}
\end{center}
\end{figure}

\begin{figure}
\begin{center}
\includegraphics[clip,width=7cm]{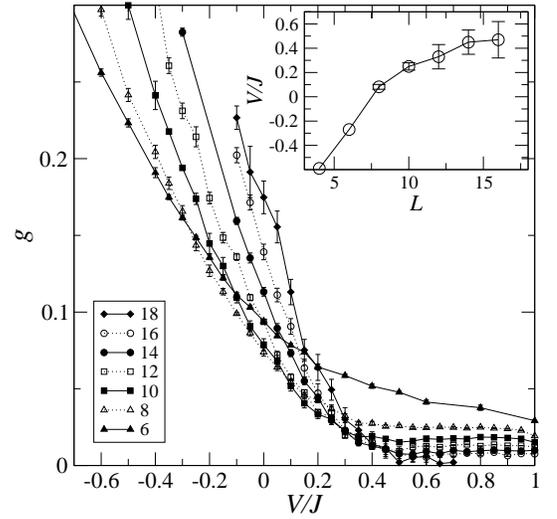}
\caption{
Binder cumulant for the order parameter $\phi$ vs. $V/J$ for different system sizes. The upper right inset is a blow-up of the region containing the crossings of the curves for the largest systems sizes.
The lower left inset shows the values of $V/J$ at intersections between $L$ and $L+2$ curves.
\label{cumulant}}
\end{center}
\end{figure}

To distinguish between these it is helpful to consider the order parameter\cite{Zhitomirsky}
\be
\phi = \f{1}{N} \sum \left( n_{\rm H}(\vec{r}) -n_{\rm V}(\vec{r}) \right)
\ee
which is unity in the columnar phase and vanishes in the plaquette phase.
A plot of $\langle \phi^2 \rangle$ as a function of $V/J$ for different system sizes is found in Fig.~\ref{phi2}.
$\phi$ is seen to decrease as $V/J$ increases. This is consistent with the plaquette phase being
more favorable as one approaches the RK-point.
In order to detect a phase transition we measure $\langle \phi^2 \rangle$
and $\langle \phi^4 \rangle$ and form the Binder cumulant
$g=1-\langle \phi^4 \rangle /(3 \langle \phi^2 \rangle ^2)$
for different system sizes. The cumulants for different system sizes should cross at the phase transition provided the scaling regime has been reached\cite{Binder}. Using exact diagonalization the Binder cumulant was calculated for system sizes up to
$L=8$ using and the phase transition was estimated to happen at roughly $V/J=-0.2$\cite{Runge}. While we reproduce the exact diagonalization results for $L=4,6,8$ we find that for larger $L$  the crossing points moves towards larger values of $V/J$, see Fig.~\ref{cumulant}.
It is rather difficult to pin down the exact location of the phase transition
based on the crossing-points summarized in the inset of Fig.~\ref{cumulant}.
However it is clear that the estimate $V/J=-0.2$ gotten from exact diagonalization studies
is too low. Simulations on larger systems are needed in order to determine the accurate value. 

Off-diagonal observables can also be measured. Consider
the operator $F$ measuring the energy associated with plaquette flips
\be
	F =  \sum \left(  \vphantom{\sum} | \verdimers \rangle \langle \hordimers | + \rm{H.c.} \right)
\ee
where the sum is to be taken over all elementary plaquettes of the lattice.
\begin{figure}
\begin{center}
\includegraphics[clip,width=7cm]{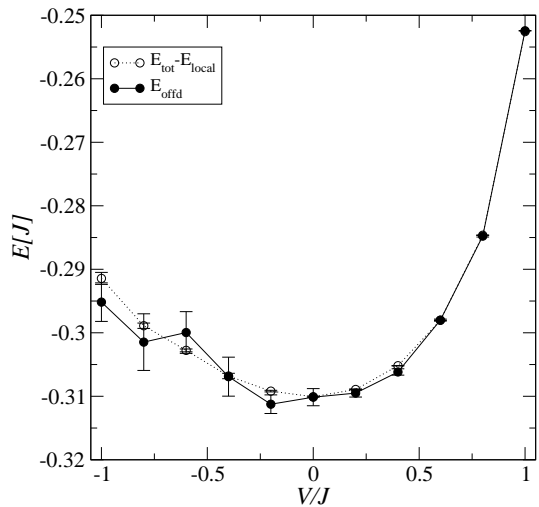}
\caption{
Offdiagonal energy associated with plaquette flips vs. $V/J$ for
a $16 \times 16$ lattice. The solid symbols denotes direct measurements of the offdiagonal terms and the open symbols are gotten from subtracting measurements
of the potential energy ($E_{local}$) from the total energy ($E_{tot}$).
\label{flipenergy}}
\end{center}
\end{figure}
Figure $\ref{flipenergy}$ shows the ground-state expectation value of
$F$ measured directly using Eq.~(\ref{eq:offdiagonal}) (solid symbols) and indirectly
measuring the potential energy which is a diagonal observable and
subtracting that from the total energy (open symbols). It is
evident that for parameters where the guiding function is
not optimal the direct measurement of the offdiagonal observable is
more noisy than the indirect measurement.

\begin{figure}
\begin{center}
    \includegraphics[clip,width=7cm]{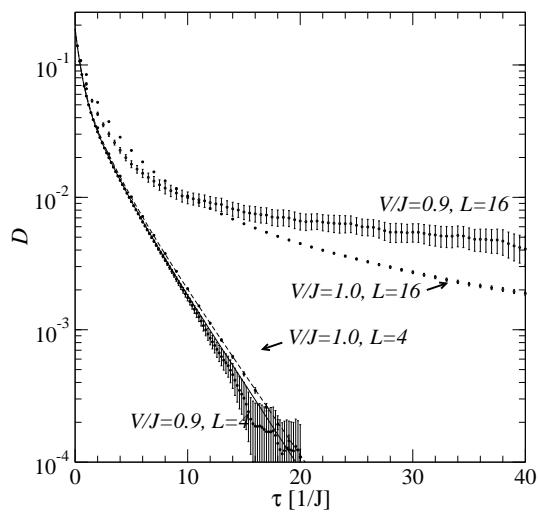}
    \caption{semi-log plot of the equal bond dimer-dimer dynamic correlation function vs. imaginary time for different values of $V/J$. The curves are for a $16\times 16$ lattice except for the two lowest curves which compares the Monte Carlo to exact diagonalization results on a $4 \times 4$ lattice.
\label{dynamics}}
\end{center}
\end{figure}

DMC has the advantage that imaginary-time correlation functions can be measured directly from the random walk.
In Fig.~\ref{dynamics} we show the equal bond dimer-dimer dynamic correlation function
$D = \langle D_i(\tau) D_i(0) \rangle - \langle D_i \rangle^2$ ($D_i$ is $1$ if a dimer is present on bond $i$ and $0$ otherwise).
The lowest two curves show the agreement with exact diagonalizations on a $4 \times 4$-lattice (lines) for two values of $V/J$. Note the semi-log
scale which ``amplifies''  the noise at small values of $D$.
 The upper two curves are for a $16 \times 16$-lattice. By fitting the long time behavior to
an exponential we find that the finite size gaps for
the $16 \times 16$-systems
are $0.02 \pm 0.01 J$ for $V/J=0.9$ and $0.022 \pm 0.001 J$ for $V/J=1.0$.

\section{Conclusions}
We have explained in details a continuous-time formulation of the DMC method. The novelty of the method
lies in its ability to output measurements of imaginary-time correlation functions without
time discretization errors. 

As any DMC method it performs best when used in combination with a good guiding function resembling 
the ground state wave function. Good guiding functions exist in the vicinity of RK-Hamiltonians, namely
the explicitly known ground state of the RK-Hamiltonian itself. Thus the method performs well for perturbed RK-Hamiltonians. 

In the case when the guiding function is equal to the exact
ground state the DMC method becomes equal to a classical Monte Carlo procedure. Thus in these
cases imaginary-time quantum correlation functions can essentially be obtained by classical
Monte Carlo. This was pointed out in Ref.~\cite{HenleyStat} for RK-Hamiltonians, but does in fact hold generally whenever the ground-state is known explicitly.

As examples of results that can be obtained using the continuous-time DMC we have
calculated order parameters away from the RK-point in the QDM on the square lattice.

%\section*{Acknowledgments}

\end{document}